# Relaxing the aquaporin crystal structure in a membrane with surface vibrational spectroscopy

L. Schmüser, M. Trefz, S. J. Roeters, W. Beckner, J. Pfaendtner, D. Otzen, S. Woutersen, M. Bonn, D. Schneider and T. Weidner

**ABSTRACT:** High-resolution structural information on membrane proteins is essential for understanding cell biology and for structure-based design of new medical drugs and drug delivery strategies. X-ray diffraction (XRD) can provide Ångstrom-level information about the structure of membrane proteins. Ideally protein structures should be solved in environments as close to the original biological context as possible. However, it is virtually impossible to crystallize proteins within the complex environment of a biological membrane. Instead, membrane proteins are typically transferred from their native membrane environment into detergent micelles, chemically stabilized and crystallized, all of which can compromise the conformation. This makes it imperative to develop alternative high-resolution techniques which are compatible with biological conditions. Here, we describe how a combination of surface-sensitive vibrational spectroscopy in model membranes and molecular dynamics simulations can account for the native membrane environment. We observe the structure of glycerol facilitator channel (GlpF), an aquaporin membrane channel finely tuned to selectively transport water and glycerol molecules across the membrane barrier. We find subtle but significant differences between the XRD structure and the inferred *in situ* structure of GlpF.

*SIGNIFICANCE: Proteins at the surfaces and interfaces of cells play important roles in biology. While methods to determine the structure of proteins with high resolution within crystals and in solution, it has been experimentally challenging to track membrane protein folding, motion and action within hydrated lipid bilayers. We show that a combination of crystallography, surface spectroscopy and computer simulations can elucidate the structure of a large membrane protein, aquaporin, when the methods are linked through theoretical modelling. Starting with the known structure of aquaporin in its crystalline form, we follow the structural relaxation into the interfacial structure within a membrane environment.*

Technologies for elucidating the atomic-level structure of membrane proteins are essential for advancing our understanding of cell biology: knowledge of membrane protein structure provides information about the transport of molecules across the membrane barrier, cell sensing, communication and also helps engineering proteins with designed functions (1).

By far the most successful methods for experimentally solving membrane protein structures with atomic resolution are nuclear magnetic resonance (NMR) (2) cryo electron microscopy (cryo EM) (3) and X-ray diffraction (XRD) (4). NMR can provide high-resolution data for proteins when incorporated in lipid vesicles and lipid stacks but is limited to smaller (typically about 20 kDa (5)) proteins and peptides. Whereas Cryo EM is restricted to larger protein with a lower size limit of 40-50 kDa. X-ray crystallography can solve the structure of very large proteins with Ångstrom resolution but requires high-quality crystals, in which proteins are removed from their native and hydrated membrane environment, and artificially stabilized. Molecular dynamics (MD) simulations can take the XRD determined structures as starting points and compensate for the non-physiological environments within crystals by relaxing the protein structure within a more 'native' biological environment.(6) The caveat with simulations has been, that the obtained structures have not been directly experimentally verified. Feedback from experimental data would be important not only to test the validity of the obtained results but also to provide feedback to improve force fields, water models, and sampling methods.

Vibrational sum frequency generation (SFG) spectroscopy is an ideal tool to directly test molecular simulations with experimental data. SFG is an inherently surface sensitive and non-invasive method, allowing the study of membrane proteins in a hydrated membrane environment (7-9). SFG makes use of the fact that vibrations of molecular groups in proteins depend strongly on both protein structure and orientation. For an SFG experiment, infrared and visible laser beams are overlapped in time and space at an interface to produce sum frequency photons of those two incident beams by nonlinear optical frequency mixing (10). Any vibrational modes in resonance with the IR beam will enhance the signal and lead to distinct spectral features. SFG spectra in the amide I region allow the analysis of the secondary and tertiary structure and orientation of proteins (11-15).

Due to the non-linear nature of SFG, spectral contributions of individual folding motifs and orientations within SFG spectra will interfere and therefore, result in complex spectra. The challenge to 'disentangle' and recover the rich structural information contained in the spectra has been the major challenge to disentangle in the SFG community for years and has recently led to protocols to calculate theoretical SFG spectra from structure files, which can be used to interpret SFG data of interfacial proteins (16, 17). Calculated SFG spectra of proteins and peptides have shown good agreement with experimental SFG spectra (18-25). However, the structural analysis of a protein the size of a membrane protein has not been reported.

Here, we take that step and observe the secondary structure of the large membrane protein aquaporin within a membrane environment by combining calculated and experimental spectra with MD simulation. In other words, we obtain the aquaporin in situ structure by experimentally verified MD simulations.

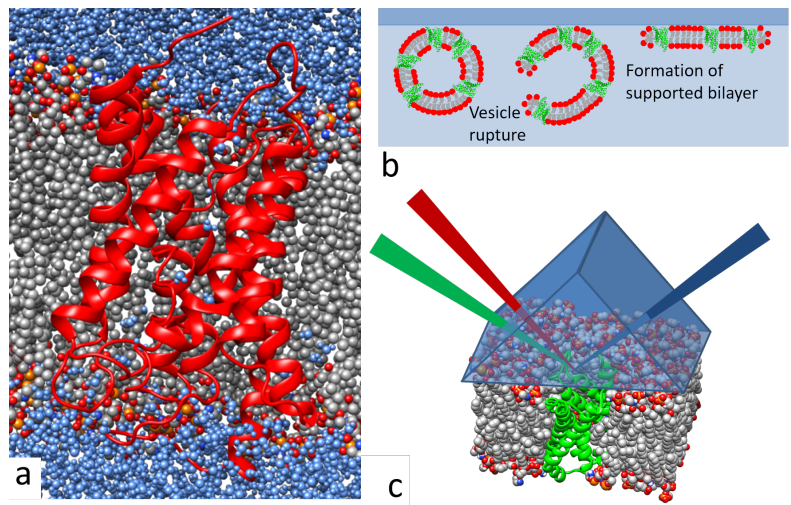

*Figure 1*. a) The Glycerol facilitator GlpF (1lda, red) embedded in a lipid bilayer (grey) with surrounding water molecules (blue), b) model of the formation of a supported lipid bilayer using liposome spreading, c) experimental setup for the SFG experiments. The lipid bilayer was formed at the solid-water-interface of an equilateral CaF$_2$ prism.

We chose an aquaporin model system because these membrane channels form an important family of membrane proteins. Water is a major component of life, and aquaporins are involved in osmoregulation of a variety of tissues in all living organisms (26). Besides facilitating a transmembrane water flux, the subfamily of aquaglyceroporins additionally facilitates the flux of glycerol and other small, polar solutes. aquaporins are found in all domains of life, e.g., 13 aquaporins are expressed in humans where they regulate the water homeostasis according to the individual requirements of the organs and cells (27). Based on their important roles in cell biology, they have great potential in diagnostics and therapeutics (28). All aquaporins form oligomers composed of four identical chains, each with one central channel.

The structure of aquaglyceroporins indicates that water (or glycerol) molecules pass through the protein channel in a single file. The aquaporin shown in Figure 1a (PDB 1lda crystallized in 28% (w/v) polyethylene glycol 2000, 100 mM Bicine, 15% (v/v) glycerol, 35 mM n-octyl-β-D-glucoside, 300 mM MgCl$_2$, and 5 mM dithiothreitol (DTT) (pH 8.9)) (29), is the well-studied glycerol facilitator channel (GlpF) of *E. coli.* A recent very high-resolution study (30) revealed how sensitively aquaporin function depends on structural fidelity: the exchange of only two amino acids at the entry to the water channel play the deciding role to ensure that water, and not hydronium or hydroxyl ions, can pass the channel (30). In our study, we now refine the XRD structure of the GlpF aquaglyceroporin shown in Fig 1a to account for the lipid environment (31).

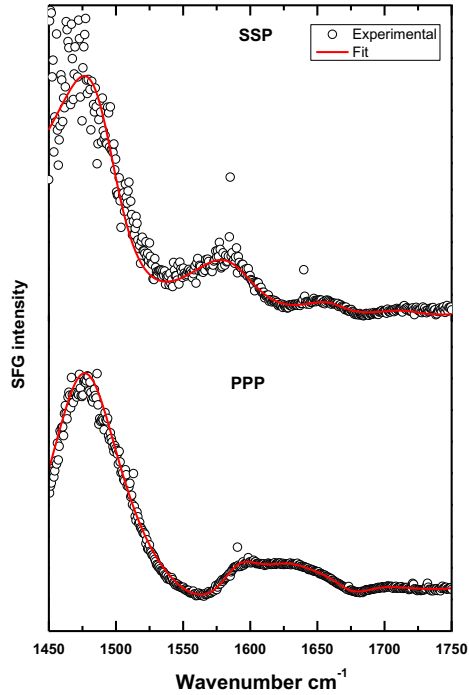

*Figure 2. Experimental SFG spectra of a GlpF including lipid bilayer, which was formed by proteoliposomes on the solid water interface of a CaF$_2$-prism mounted on a flow cell. The two different laser polarization combinations SSP and PPP have been measured. The spectra have been fitted (red line) using a Lorentzian lineshape model (Eq1 and Eq2).*

We spread GlpF-loaded proteoliposomes, prepared with *E. coli* polar lipid extract (see SI), on one side of a CaF$_2$ prism (Figure 1b). The liposomes spread at the surface to form a supported lipid bilayer, which was then probed through the backside of the prism with SFG in near total internal reflection geometry (Figure 1c). Representative experimental spectra in SSP (S-polarized SFG, S-polarized visible and P-polarized infrared) and PPP polarization are shown in Fig. 2. The observed spectra are typical for the entangled, complex signal expected from large protein systems. Nonetheless, the spectrum can be described using the following standard equation describing the SFG response from an interface (32):

$$I_{SFG}(\omega_{IR}) \propto \left|(\chi^{(2)}\omega_{IR})\right|^2 I_{vis} I_{IR} \qquad (1)$$

$$\chi^{(2)}\omega_{IR} = A_0 e^{i\varphi} + \sum_n [A_n / (\omega_{IR} - \omega_n + i\Gamma_n)] \qquad (2)$$

Here, $I_{SFG}(\omega_{IR})$, $I_{vis} I_{IR}$, $\chi^{(2)}\omega_{IR}$, $A_0$, and $\varphi$ are the SFG intensity, the intensities of IR and visible laser beams, the second order susceptibility, the amplitude and phase of the nonresonant background. The spectra are dominated by a peak near 1480 cm$^{-1}$, which can be attributed to CH$_2$ and CH$_3$ scissoring modes of both lipids and protein (33). A feature near 1650 cm$^{-1}$ is representative of the α-helical structure of the membrane protein (33). In addition, there is a smaller peak near 1720 cm$^{-1}$, which originates from carbonyl groups of membrane lipids. The feature observed near 1600 cm$^{-1}$ is not well described in literature and may represent a new structural feature.

To directly relate the SFG measurements to protein structure, we first calculate the SFG spectra from GlpF structure files obtained with XRD from protein crystals (PDB 1lda) (29). The calculation followed the procedure described by Roeters et al. (16) The model takes into account all CO, CN, NH and CC bond vibrations along the peptide backbone (see SI). The theoretical spectra for the protein crystals are shown in Figure 3 (Purple dashed line). To compare with the experimental protein backbone data, we plot the backbone related components of the fit to the experimental data in Figure 3 (solid line) and exclude any non-protein modes from the experimental data, which could obscure the protein peaks. While the calculated spectra from the XRD structure correctly predict intensity near 1650 cm$^{-1}$ for both ssp and ppp polarization combinations, the feature near 1600 cm$^{-1}$ is not captured correctly by the calculations. The discrepancy between the calculated and observed spectra indicates that the structure of GlpF within the stabilized protein crystal used for XRD differs substantially from its structure when incorporated in a lipid bilayer.

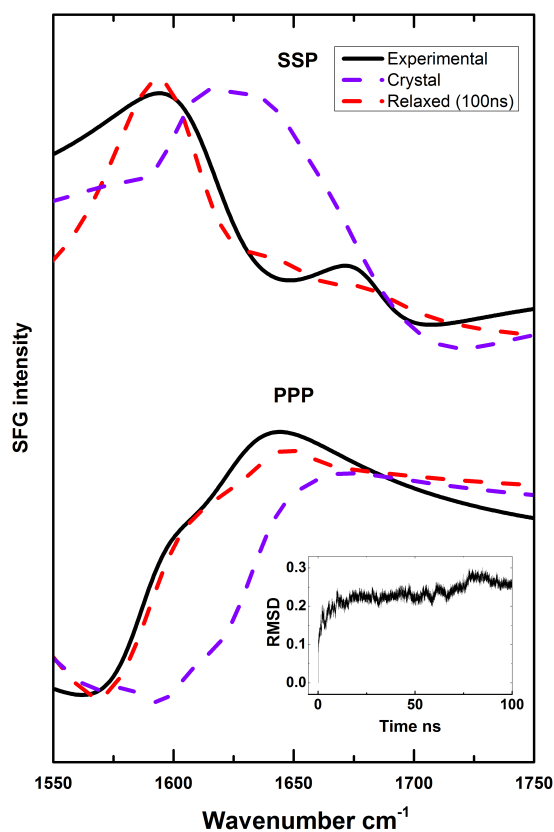

*Figure 3. comparison of protein specific SFG spectra extracted from fits of experimental data (solid lines) and calculated SFG spectra of the MD simulation after 100ns (dashed lines). The inlet represents the root-mean-square deviation of atomic positions plotted against the simulation time showing that a stable conformation was already reached after a few ns.*

This discrepancy between the observed and expected SFG response can be resolved when going from the crystalline state represented by the XRD structure to the native lipid membrane state. Using the XRD structure as the starting point, we ran MD simulations (see SI) of GlpF within a 1-Palmitoyl-2-oleoyl-sn-glycero-3-phosphoethanolamine (POPE) bilayer until the structure was equilibrated (~100 ns, see Figure 3). Indeed, significant changes in the structure were apparent.

Spectra calculated from this 'relaxed' structure yielded theoretical spectra with a substantially improved match to the measured SFG spectra (Figure 3 and S3, red dashed line). The spectra based on the simulated model capture the spectral shape of the experimental data very well. The 1600 cm$^{-1}$ peak and also the intensity at higher energies now match much better compared with the spectrum calculated using the crystal structure. The root mean squared deviation (RMSD) of the computed spectra from the experimental data decreased by 55% for the relaxed structures compared with the crystal structure (see SI). The RMSD values between relaxed structures taken at different time points of the simulation varied only by approximately 5-10% (see Figure S4). A mode analysis shows that the modes contributing to the signal are evenly distributed across the protein (see Figure S2), and any changes to the calculated spectra are a result of changes within the global protein structure. Note that, apart from an overall amplitude scaling factor, there are no adjustable parameters when comparing the experimental and theoretical results.

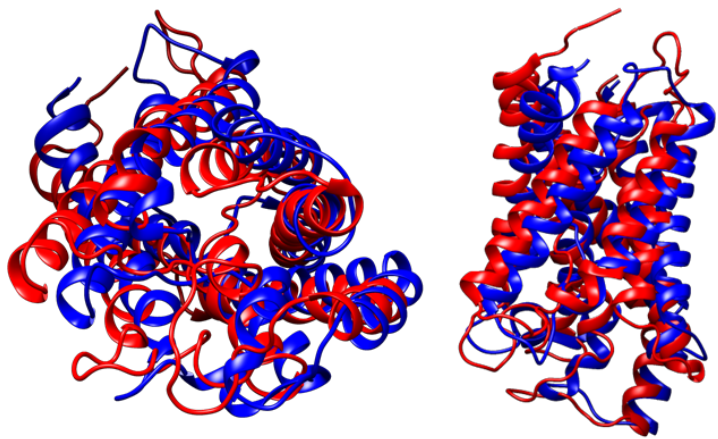

*Figure 4. Structural differences between the crystal structure (1lda, blue) and the refined protein structure (red). The refined structure was obtained from an MD simulation after 100 ns and directly validated by SFG.*

Figure 4 shows the crystal-based GlpF structure along with the refined structure that is consistent with the SFG results. There are marked differences between the two structures: both in the loop regions, and particularly in the helix orientations there are variations. It is well known that aquaporin action depends on such minute structural details (34). Also, the loop regions appear to be crucial for proper activity (35). Knowledge of structural details is therefore crucial for interpreting the changes in the substrate conducting mechanism induced, e.g. by mutations in aqua(glycero)porins and in membrane proteins in general. The SFG response is very sensitive to the fairly subtle changes in the protein structure shown in Fig. 4, likely because of changes in the arrangement of the helices.

The method described here combines the resolving power of MD simulations and the interface and structure sensitivity of SFG. The resulting hybrid experimental and theoretical structures allow adjustments upon existing XRD protein crystal data to include the influence of the hydrated lipid bilayer. Moreover, this method can also determine the impact of therapeutic drugs on aquaporins in situ. As such, we expect the combination of MD simulations and SFG to be an asset to future studies of selective molecular transport within aquaporins, and the function of a broad variety of membrane proteins.

## ASSOCIATED CONTENT

**Supporting Information**. Details of the SFG experiments and data analysis as well as the simulations and the spectra calculations are summarized. The Supporting Information is available as a PDF free of charge..

## AUTHOR INFORMATION

### Author contributions

L.S., T.W., J.P. and D.S. designed the experiment. L.S. performed the SFG experiments. M.T. isolated and prepared the proteoliposomes. S.J.R and W.B. performed the calculations and simulation. J.P., S.W., M.B., D.S. and T.W. analyzed data. All authors discussed the results and wrote the manuscript.

### Corresponding Author

Tobias Weidner, weidner@chem.au.dk

## ACKNOWLEDGMENT

This work was supported by the Max Planck Graduate Center with the Johannes Gutenberg-Universität Mainz. SJR and TW thank the Villum Foundation (Experiment Grant 22956) for financial support. Partial financial support to JP and WB was provided by NSF (CBET-1264459).

# Supporting information:
# Relaxing the aquaporin crystal structure in a membrane with surface vibrational spectroscopy

L. Schmüser, M. Trefz, S. J. Roeters, W. Beckner, J. Pfaendtner, D. Otzen, S. Woutersen, M. Bonn, D. Schneider and T. Weidner

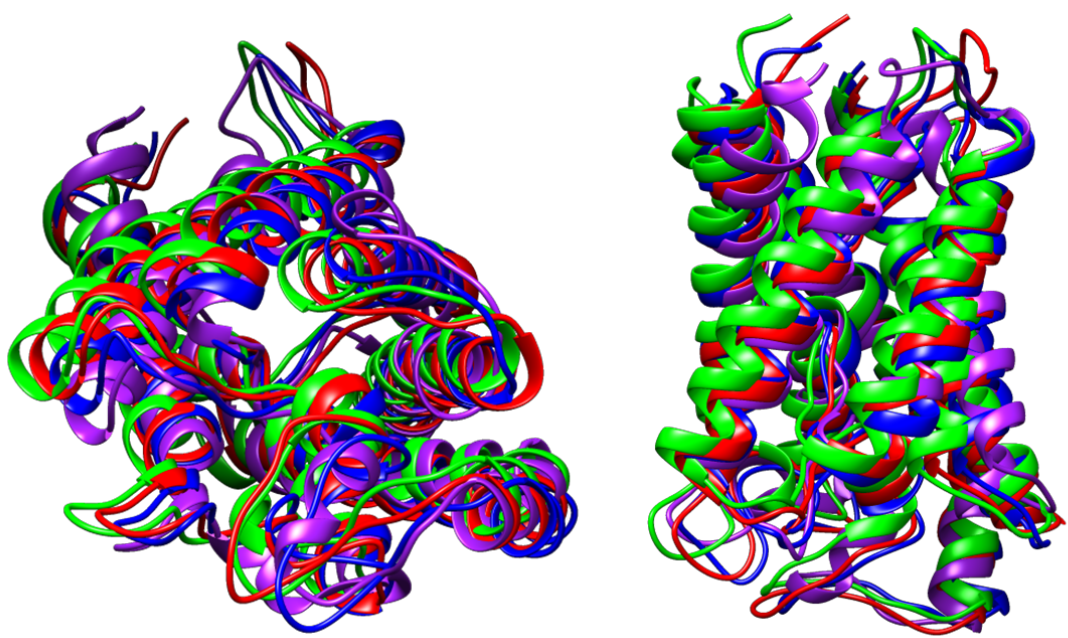

**Figure S1.** Comparison between the crystal structure (1lda, purple) and MD snapshots after 100ns (red), 75ns (green) and 50ns (blue). The structural overlays were created using the chimera tool matchmaker. Relaxing the protein with MD simulation changed slightly the tilt angles of its alpha helical elements.

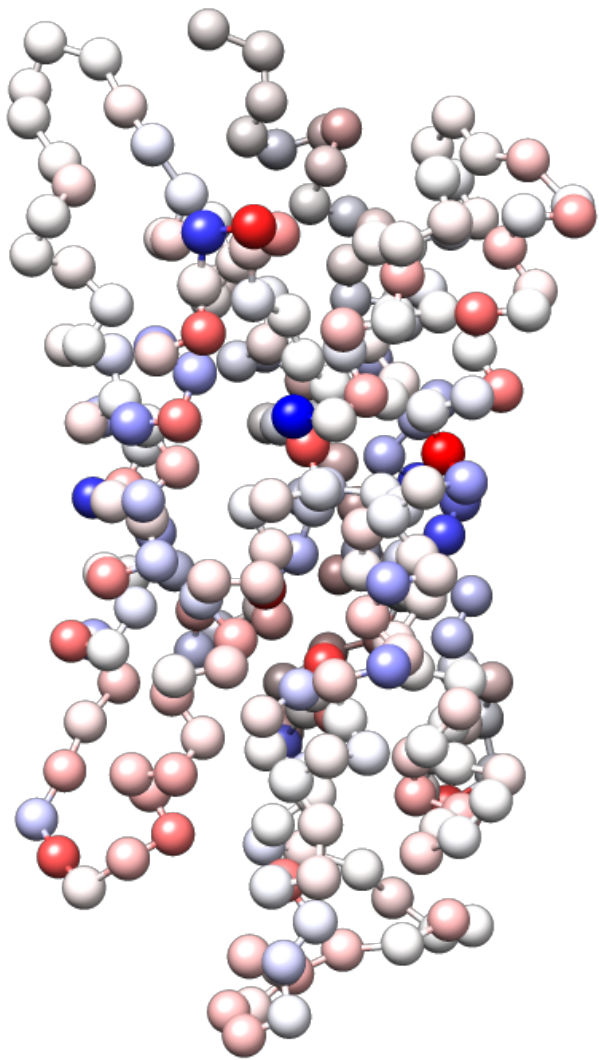

**Figure S2**. Averaged contributions of amino acids to the eigenmodes around 1600 cm$^{-1}$ proving that the origin of the main peak at 1600 cm$^{-1}$ is broadly distributed over the whole protein. Blue and red spheres are indicating a high contribution with a negative (blue) or positive (red) amplitude. This figure is plotted using Chimera (UCSF). The amplitudes for each eigenmode are derived from the Mathematica script, which calculates the SFG spectra.

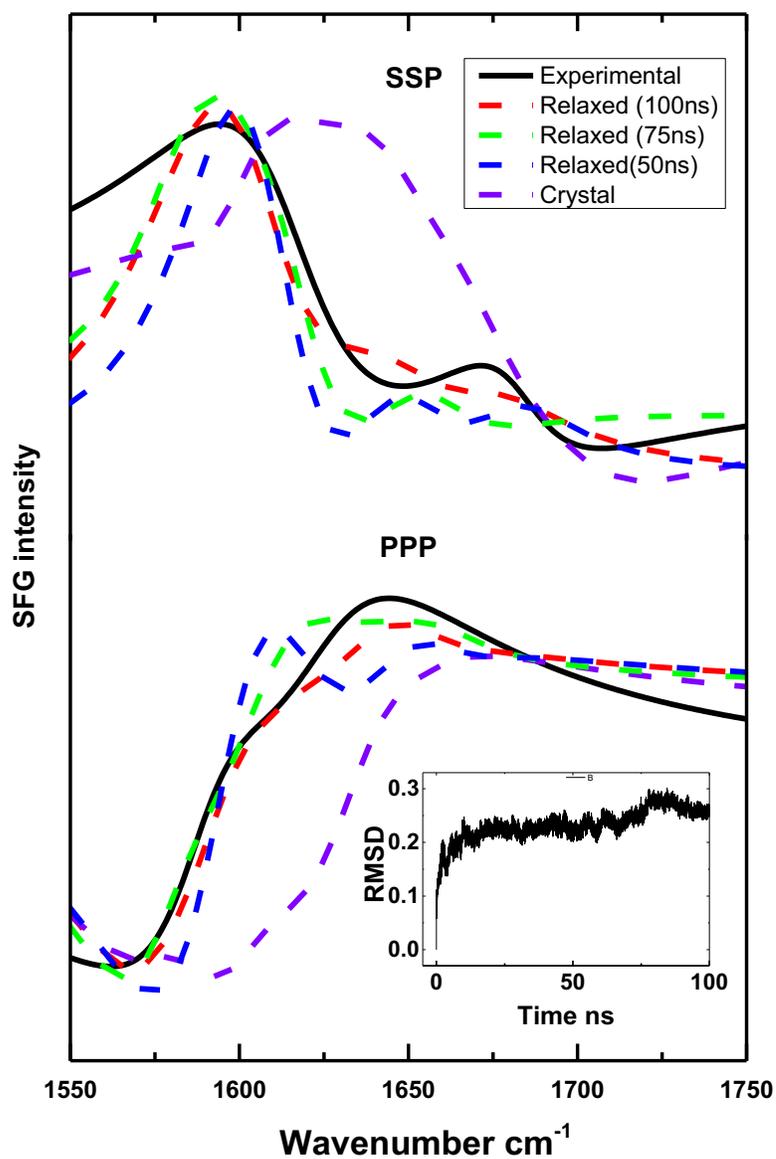

**Figure S3.** Comparison of the experimental SFG spectra (black solid line) with calculated SFG spectra of the crystal structure (dashed purple line) and of different snapshots of the MD simulation (green 100 ns, red 75 ns, and blue 50 ns, see S1). The inlet represents the root mean square deviation of atomic distances between the starting structure (the crystal structure) and structures of the MD simulation, plotted against the simulation time.

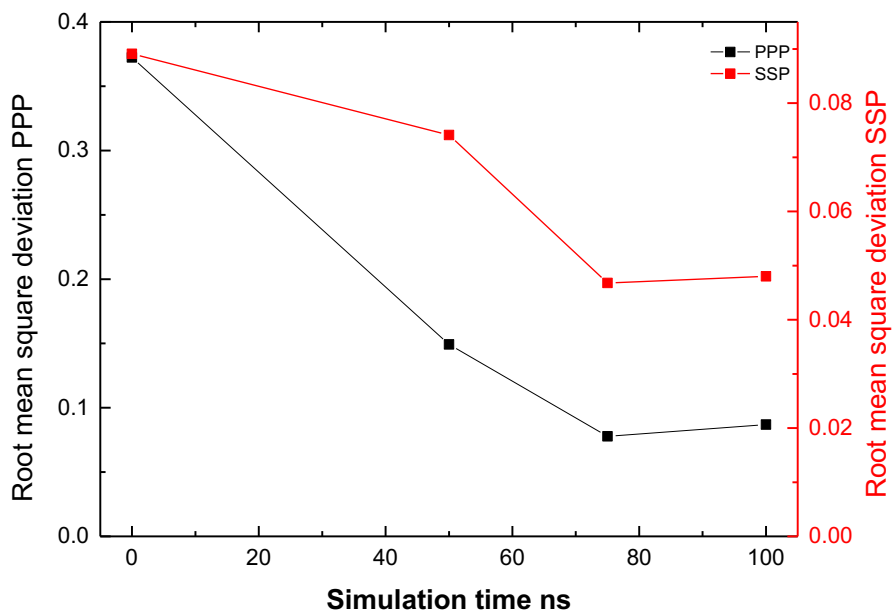

**Figure S4.** Quantification of the structural agreement between the aquaporin GlpF in physiological condition with the X-ray crystallography structure (PDB: 1lda) and with the structures derived from MD simulations. The differences between the experimental SFG spectra and those which were calculated from the PDB file 1lda and the MD simulations were determined using the root mean square deviations (RMSD) between experimental and calculated SFG spectra. The RMSD was calculated using the following equation:

$$RMSD = \sqrt{\sum_{\omega=1550}^{1750} \frac{(I\ exp_\omega - I\ calc_\omega)^2}{200}}$$

Here, $\omega$ is the wavenumber and *I exp$_\omega$* and *I calc$_\omega$* are the experimental and calculated SFG intensities respectively. The calculated SFG spectra in Fig S3 were scaled to match the experimental spectra.

## Methods

### Expression and purification of GlpF into proteoliposomes

GlpF expression and purification were performed as described in detail in(1). The concentration of purified GlpF was determined by absorption measurements at 280 nm, using a calculated extinction coefficient of 37930 $M^{-1}$ $cm^{-1}$ (ExPasy ProtParam tool).

Liposomes prepared from *E. coli* polar lipid extract (EPL: 67% PE, 23.2% PG and 9.8% CL) and 1,2-dioleoyl-sn-glycero-3-phosphocholine (diC18:1-PC, DOPC) were used for GlpF reconstitution(2). Chloroform lipid solutions were purchased from Avanti Polar Lipids (Alabaster, AL). To remove the organic solvent, 5 mM of the dissolved lipid was set under a stream of nitrogen gas. Remaining organic solvent was removed by overnight vacuum desiccation. The resulting lipid film was rehydrated in 30 mM n-octyl β-D-glycopyranoside (OG; Roth, Karlsruhe, Germany), 50 mM MOPS pH 7.5 (Sigma-Aldrich, Munich, Germany), 150 mM N-methyl D-glucamine (Acros Organics, Morris Plains, NJ) and 50 mM NaCl (Roth, Karlsruhe, Germany) at 37 °C for 45 min. Purified GlpF was added to the rehydrated lipids to reach a final lipid concentration of 12 µM and a molar lipid/GlpF ratio of 400:1. The final sample was adjusted to a volume of 0.5 mL and an OG concentration of 30 mM by addition of buffer (50 mM MOPS, pH 7.5, 150 mM N-methyl D-glucamine, 50 mM NaCl). Subsequently, the sample was dialyzed for 48 h at 4 °C against 500 mL MOPS buffer. The dialysis buffer was exchanged three times.

### SFG experiment

The SFG setup has been described in detail before.(3) Briefly a 40 fs, 5 mJ, 800 nm visible pulse was generated by a regenerative amplifier (Spitfire ACE, Spectra Physics) using a Nd:YLF pulse laser (Empower, Spectra Physics) and a Ti:sapphire seed laser (MaiTai, Spectra Physics). One part of the 800 nm beam was branched out to pump an optical parametric amplifier (TOPAS, Spectra Physics), which generates a 40 fs broadband IR pulse. The remaining 800 nm beam was spectrally narrowed to a 25 µJ 15 $cm^{-1}$ FWHM pulse using a Fabry-Perot Etalon (SLS Optics Ltd.) and temporally and spatially overlapped with the IR pulse. IR and visible laser pulses were both focused on the sample. The laser polarization combinations SSP(S-polarized SFG, S-polarized visible and P-polarized IR) and PPP were obtained using polarizers and half-wave-plates in each beam path. The generated SFG signal was collimated using lenses and separated from the visible light using low pass filters. The focused SFG signal was directed onto a spectrograph (Acton Instruments) and finally detected by a CCD camera (Newton, Andor Technologies).

The SFG experiments were performed in a nitrogen-flushed chamber to avoid absorption of the IR pulse due to water vapor. The experiments were done in a flow cell with a volume of 1 ml. The flow cell was sealed on one side with an equilateral $CaF_2$ Prism. 400 µl of proteoliposome solution was injected into the flow cell and incubated for at least 2 h. Remaining proteoliposomes in bulk were rinsed with $D_2O$ followed by an overnight waiting step to allow hydrogen to deuterium exchange. SFG spectra were collected in SSP and PPP polarization combination and normalized using reference spectra of a $CaF_2$ prism, which was coated with a 100 nm silver film at the $CaF_2$ water interface.

**Theory of calculation**

First we obtain the atom coordinates of the amide groups from the PDB file of the crystal structure or of the MD simulation. We use the coordinates both to calculate the transition dipole moments of the local modes (by determining the transition charge of each atom(4), as this gives more accurate spectra than the conventional approach as well as the Raman polarizabilities similar to refs.(5, 6), and to construct the one-exciton Hamiltonian:

$$H = \begin{pmatrix} \hbar\omega^0_1 & \kappa_{12} & \kappa_{13} & \kappa_{14} & L \\ \kappa_{12} & \hbar\omega^0_2 & \kappa_{23} & \kappa_{24} & \\ \kappa_{13} & \kappa_{23} & \hbar\omega^0_3 & \kappa_{34} & \\ \kappa_{14} & \kappa_{24} & \kappa_{34} & \hbar\omega^0_4 & \\ M & & & & O \end{pmatrix}$$

(3)

with $\omega^0_i$ the gas phase frequency of local mode $i$ and $\kappa_{ij}$ the coupling between local mode $i$ and $j$.

The diagonal terms are determined with using an empirical model that gives the local mode frequency as a function of the strength of the three possible hydrogen bonds that each amide group can form(7), comparable to the model used in ref.(8).

For the off-diagonal terms, the couplings between the normal modes, we discriminate between nearest-neighbor and non-nearest-neighbor coupling. As the former is dominated by through bond charge flows we use a parameterization of the coupling as a function of the two dihedral angles between the two neighboring amide groups calculated for the "glycine dipeptide" (Ac-Gly-NHCH3), using the 6-31G+(d) basis set and B3LYP-functional(9).

The non- nearest-neighbor couplings are dominated by through-space (Coulomb) interactions, so we estimate these with the Transition Dipole Coupling method(10):

$$\kappa_{ij} = \frac{1}{4\pi\varepsilon_0}\left(\frac{\vec{\mu}_i \cdot \vec{\mu}_j}{|\vec{r}_{ij}|^3} - 3\frac{(\vec{r}_{ij} \cdot \vec{\mu}_i)(\vec{r}_{ij} \cdot \vec{\mu}_j)}{|\vec{r}_{ij}|^5}\right) \quad (4)$$

with $\vec{\mu}_i$ the transition dipole moment of local mode $i$, $\vec{r}_{ij}$ the distance between local mode $i$ and $j$, and $\varepsilon_0$ the dielectric constant.

Subsequently the Hamiltonian is diagonalized to obtain the normal mode eigenvalues and eigenvectors, from which the IR, Raman and VSFG responses are calculated, according to ref. (7) in which also other details regarding the formalism used here for the spectral calculations can be found.

The non-resonant phase and its amplitude were adapted to yield the best match between the experimental data and the calculated ones. Otherwise, all calculations were done using the same settings.

**MD simulation**

Simulations of the porin tetramer (1lda(11)) and phosphatidylethanolamine (POPE) lipid bilayer were performed using the GROMACS(12) molecular dynamics (MD) engine and explicitly solvated using the TIP3P(13) water model. Force fields for the lipid bilayer were taken from Tieleman and Berendsen(14) and adapted into the GROMOS96 53A6 force field(15), which is extended to include Berger lipid parameters and has been verified to perform as well or better than previous versions for protein simulations(16). 2.0 fs time steps in the MD simulation were integrated using a leap-frog algorithm(17). For distances exceeding 1.2 nm, van der Waals interactions were shifted to 0 with a switching function applied at 1.0 nm and electrostatic forces were treated with particle mesh Ewald (PME) summation. Bond lengths between Hydrogen and heavy atoms were fixed with the LINCS linear constraint solver(18).

The tetramer and lipid bilayer were centered in an overall charge-neutral system with periodic x, y, z dimensions of 11.49, 11.38, and 10.39 nm, respectively. The system was restricted from lateral-diffusion of the membranes by restraining the relative motion of the protein and bilayer to the solvent. Following energy minimization, a 100 ps NVT simulation was conducted at a temperature above the phase transition temperature (T = 298 K) of the lipid membrane(19) to allow the equilibration of water and ions. Protein, lipid, and solvent/ions were temperature coupled independently at 315 K using a stochastic global thermostat(20) and a 0.1 ps coupling constant.

A 1 ns NPT equilibration step was conducted after NVT equilibration. The thermostat was switched to Nose-Hoover(21) with a 0.4 ps coupling constant to more realistically capture temperature fluctuations(22). Semi-isotropic pressure coupling was used to allow the membrane to deform in the xy plane independently of z. Following NVT and NPT equilibration, position restraints on the tetramer and lipid bilayer were relaxed and the system underwent 100 ns of production MD in the NPT ensemble. Coordinates of the resulting structure were used for SFG analysis from time point 100 ns.